\documentclass{article}
\usepackage{mn2e}
\usepackage{mc852-bib}
\usepackage{epsfig}
\begin{document}
\newcommand{\pI}{Paper I}
\newcommand{\pII}{Paper II}
\newcommand{\pIandII}{Papers I and II}

\title[Line Indices of PG\,1605+072]
{Time-series Spectroscopy of Pulsating sdB Stars III:
  Line Indices of PG\,1605+072\thanks{Based on observations made
    with the Danish 1.54\,m telescope at ESO, La Silla, Chile. Part of
    the data presented here have been taken using ALFOSC, which is
    owned by the Instituto de Astrofisica de Andalucia (IAA) and
    operated at the Nordic Optical Telescope under agreement between
    IAA and the NBIfAFG of the Astronomical Observatory of Copenhagen.}}

\author[S.~J.~O'Toole et al.]{S.~J.~O'Toole,$^1$
  M.~S.~J{\o}rgensen,$^2$ H.~Kjeldsen,$^{3,4}$ T.~R.~Bedding,$^1$
  T.~H.~Dall,$^4$\cr and U.~Heber.$^5$ \\
$^1$School of Physics, University of Sydney, NSW 2006, Australia\\
$^2$University of Copenhagen, Astronomical Observatory, Juliane Maries
  Vej 30, DK-2100 Copenhagen \O, Denmark\\
$^3$Theoretical Astrophysics Center, Aarhus University, DK-8000,
  Aarhus C, Denmark\\
$^4$Institute of Physics and Astronomy, Aarhus University, DK-8000,
  Aarhus C, Denmark\\
$^5$Dr Remeis-Sternwarte, Universit\"at Erlangen-N\"urnberg,
  Sternwartstr. 7, D-96049 Bamberg, Germany}

\maketitle

\begin{abstract}

We present the detection and analysis of line index variations
in the pulsating sdB star PG\,1605+072. We have found a strong
dependence of line index amplitude on Balmer line order, with
high-order Balmer line amplitudes up to 10 times larger than
H$\beta$. Using a simple model, we have found that the line index
may not only be dependent on temperature, as is usually assumed for
oscillating stars, but also on surface gravity. This information will
provide another set of observables that can be used for mode
identification of sdBs.

\end{abstract}


\section{Introduction}
\label{sec:intro-ew}

Subdwarf B (sdB) stars are hot, core-helium burning stars, with hydrogen
envelopes that are too thin to sustain nuclear fusion. The discovery
of multimode pulsations in some sdBs should allow the use of
asteroseismology to probe their atmospheres, and thereby help to
answer the many questions remaining about sdB formation and evolution.
 
Until recently, time-series observations of pulsating sdBs have been
limited to photometry. We have shown that it is possible to detect
radial velocity variations of Balmer lines in the pulsating sdB star
PG\,1605+072 using 2\,m-class telescopes (\citetwobare{OBK00b}{OBK02},
hereafter \pIandII). Four-metre class telescopes have also been used
to detect velocity variations in sdBs \cite{JP00,WJP01}. In \pII\ we
found closely spaced peaks in the amplitude spectrum of
PG\,1605+072, as well as amplitude variation in at least one of these
peaks. In this paper we describe the analysis of
line-index variations of Balmer lines in PG\,1605+072. The
atmospheric parameters of this sdB have been determined from
high-resolution optical spectra and NLTE model atmospheres to be
$T_\mathrm{eff}$=32\,300$\pm$300\,K and log\,$g$=5.25$\pm$0.05
\cite{HRW99}. The helium abundance was found to be subsolar, i.e.,
log\,(He/H)=$-$2.53$\pm$0.1. The star was also found to be rotating
with a projected rotational velocity of 39\,km\,s$^{-1}$.

The use of equivalent widths of Balmer lines to monitor stellar
oscillations of solar-type stars was first proposed by
\citeone{KBV95}. The Balmer lines are very temperature sensitive --
the equivalent width of these lines depends on the number of hydrogen
atoms in the second level. According to \citeone{Kurucz79}, the
maximum line strength of H$\alpha$ to H$\delta$ is between 8500\,K and
9000\,K at log\,$g$=4.0. Since the temperature
changes slightly with the oscillations, the equivalent width must also
change. The idea was extended by \citeone{BKR96} to include metal
lines. In this paper we calculate \textit{line indices}, which depend
on the temperature and surface gravity sensitivity of the equivalent
widths of spectral lines. This method has been used to identify modes
in the $\delta$ Scuti stars FG Vir \cite{VKB98} and BN Cnc
\cite{DFL02}, and to detect equivalent-width oscillations in the
H$\alpha$ line of the roAp star $\alpha$ Cir \cite{BVB99}. Here we
present the first detection of line index variations in a pulsating
sdB star, namely PG\,1605+072, and find a dramatic dependence of
line-index amplitude on Balmer line number. We show that a simple
model can qualitatively explain this effect as a combination of
temperature and surface gravity fluctuations.

\section{Observations and Reductions}
\label{sec:obsred}

In this paper we use the spectroscopic observations described in
\pIandII. Briefly, they consist of 16 nights in May 2000 using DFOSC
on the Danish 1.54\,m in Chile, 4 nights overlapping this using ALFOSC
on the NOT 2.56\,m and, to improve frequency resolution, a further
$\sim$1 hour per night for a month in March--April 2000 using
DFOSC. We also have 10 nights of observations using DFOSC from July
1999 using the same setup, as well as 2 nights using the
coud\'e spectrograph mounted on the Mount Stromlo 74-inch telescope in
Australia. Finally, we have 12 nights of Johnson $B$ photometry from
the South African Astronomical Observatory, also taken in July 1999
(see \pI).

\begin{figure}
  \begin{center}
    \leavevmode
    \epsfig{file=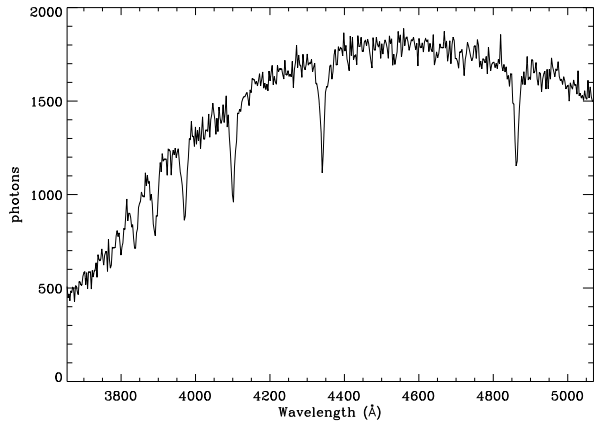,scale=0.47}
    \caption{Spectrum of PG\,1605+072 taken using DFOSC at the Danish
    1.54\,m telescope. The continuum is clear until around H$\epsilon$.}
    \label{fig:showspec}
  \end{center}
\end{figure}

As described in \pIandII, reductions of the spectra were done using
standard IRAF routines for bias subtraction, flat fielding and
background light subtraction, and spectra were extracted using a
variance weighting algorithm. A typical spectrum is shown in Figure
\ref{fig:showspec}. For details of photometric reductions
see, for example, \citeone{ECpaperII}.

\section{The Line Index}
\label{sec:Lamdesc}

We measure the line index, a quantity which depends on the
temperature and surface gravity sensitivity of the equivalent
width. It is defined as
\begin{equation}
\Lambda^\mathrm{L}=\sum_x\frac{C_x-S_x}{C_x}W_x^\mathrm{L}
\label{eq:LI}
\end{equation}
\cite{DFL02}, where $C_x$ is the continuum count level at pixel $x$,
$S_x$ is the source count level, and $W_x^L$ is a gaussian-like
weighting functionfor the line in question. The value of $C_x$ is
found using $C_x=\sum S_xW_x^\mathrm{C}/\sum W_x^\mathrm{C}$ where
$W_x^\mathrm{C}$ is a broad filter function and is also
gaussian-like. $W_x^\mathrm{L}$ is centered by moving an integration
weight filter across the line in question. The position that maximises
the sum through this filter is taken as the line centre
\cite{DFL02}. The width of the function can be chosen arbitrarily; we
have chosen the approximate full width at half maximum of each Balmer
line, which means we use a different filter for each line.

Prior to calculating the line index, each spectrum was normalised to
an average number of counts per pixel and divided by the
local continuum. The line index was calculated by multiplying each
Balmer line and the adjacent continuum by the weighting function
(a super-super-gaussian: $e^{-x^8}$), dividing the line by the average of
the two continua, and then summing over the line. In this way, the
line index can be thought of as similar to the Str\"omgren H$\beta$
index, calculated using software filters. (For a justification of the
term ``line index'' as an analogue to a colour index, see
\citebare{DFL02}). The software for calculating line indices has been
collected in a software package called \textit{Ix}, as described by
\citeone{Dall2000}.

\subsection{Individual Balmer Lines}
\label{sec:balmer-ew}

\begin{figure}
\center{\epsfig{file=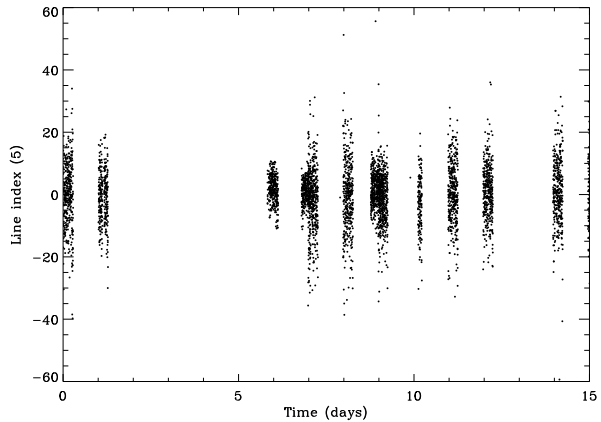,scale=0.47}}
\caption{Line index as a function of time for the H8 line.}
\label{fig:EWcurve}
\end{figure}

\begin{figure*}
\epsfig{file=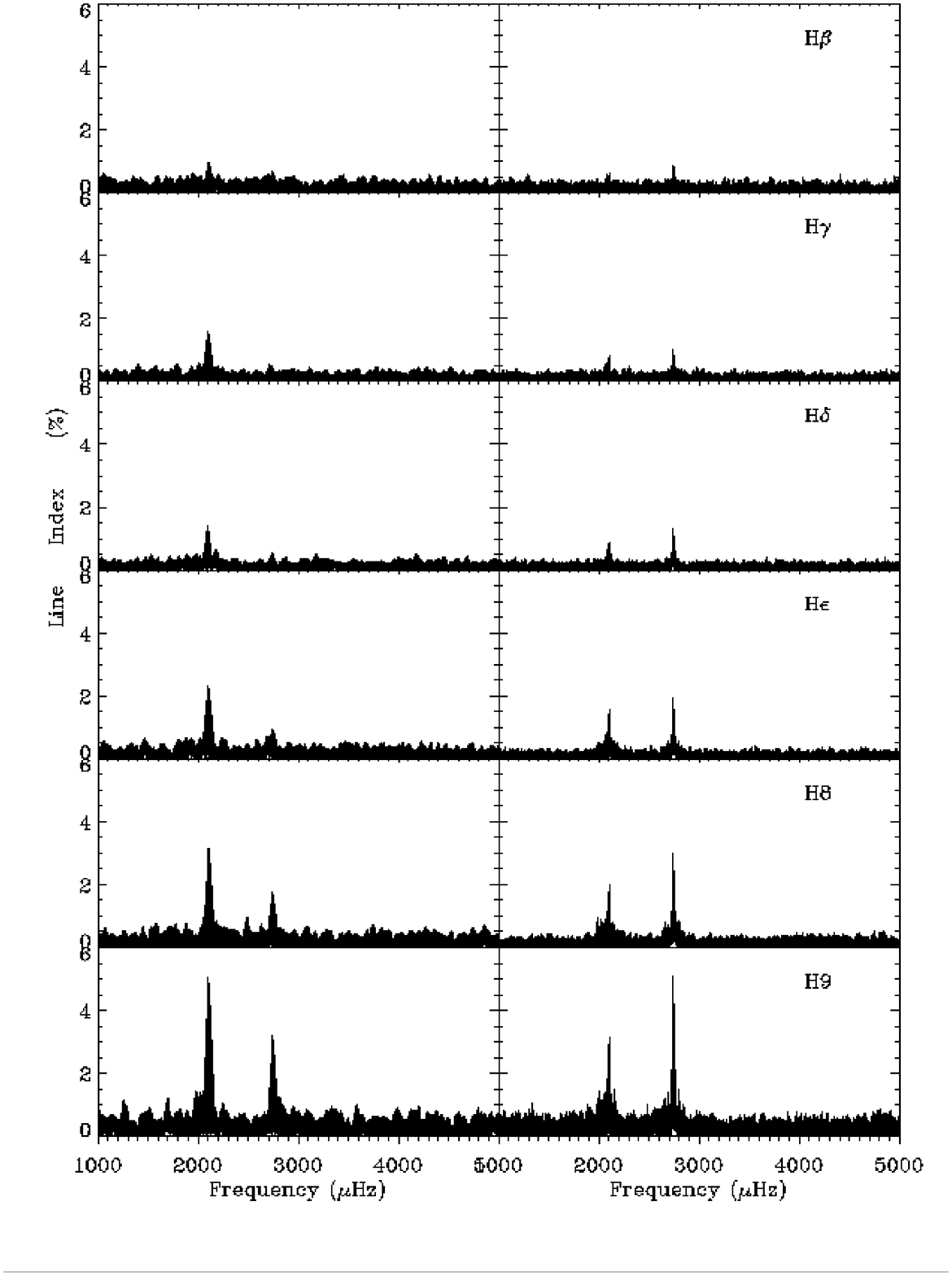,scale=0.85}
\caption{Line index amplitude spectra for 6 Balmer lines,
  from H$\beta$ at the top to H9 at the bottom. The left panels are
  based on July 1999 observations and the right panels on March-May
  2000 observations. A dramatic increase in amplitude is evident
  towards higher-order Balmer lines.}
\label{fig:ewspec}
\end{figure*}

We have calculated the line indices of all Balmer lines from
H$\beta$ to H9. These values were normalised to a mean of zero and
divided by the total mean to give the fractional change (i.e. we
calculated
\mbox{$(\Lambda-\langle\Lambda\rangle)/\langle\Lambda\rangle$}, where
$\langle\Lambda\rangle$ is the mean line index). A sample
line index curve (of H8) is shown in Figure \ref{fig:EWcurve};
the peak-to-peak scatter of the best quality data (from the NOT) is
around 20\%. As in \pIandII, the quality of the data varies from night
to night, so we have weighted the Fourier transform using the internal
scatter. 

The amplitude spectra for the six Balmer lines from H$\beta$ to
H9 is shown in Figure \ref{fig:ewspec}. The left column shows results
for the 1999 data (\pI), while the right column shows the 2000 data
(\pII). An amplitude change is evident over the 300 days between
observations, similar to that seen in velocity. Also, line-index
amplitudes vary strongly with Balmer line number in both sets of
observations. We now consider possible reasons for the latter effect.

\subsection{Dependence of amplitude of Balmer-line order}
\label{sec:explain}

Could the variation with Balmer line order seen in Figure
\ref{fig:ewspec} be an artifact of our measurement technique? There
are several factors we should consider.

Firstly, in sdB stars the Balmer lines are broadened by
temperature and gravity, so it is difficult to measure the continuum
near high-order lines ($n$=9--12), since it does not really exist in
these regions. However, the amplitudes of the strongest peaks
differ by about a factor of three between H$\beta$ and H$\epsilon$,
and these lines have good continuum, as seen in Figure
\ref{fig:showspec}. Still, the wings of the Balmer lines are very
broad in sdBs and are dominated by gravity effects, so we will
investigate this ``pseudo-continuum'' at high order Balmer lines in
Section \ref{sec:simpletest}.

A second explanation may be that the wings of the Balmer lines are
oscillating out of phase with the cores. Since the wings extend out so
far, they may be inadvertantly sampled as part of the continuum and
thereby boost the observed amplitude of line index
change. We have determined the line index variation
across each Balmer line at each of the
frequencies in Table 7 of \pII. The phases at each frequency are
constant when we examine different slices across the line, leading us
to rule out this explanation.

\section{A Simple Model}
\label{sec:simpletest}

To investigate the observed behaviour, we have used H-He line
blanketed, metal-free NLTE model sdB spectra \cite{Napi97} at
log\,$g$=5.0, 5.25, 5.5 and $T_{\mathrm{eff}}$=30\,000\,K, 32\,500\,K
and 35\,000\,K. The helium abundance is fixed at
log\,(He/H)$=-2.5$. The calculations that follow were also carried out
using metal line-blanketed LTE model sdB spectra with solar
metallicity and Kurucz' ATLAS6 Opacity Distribution Functions
\cite{HRW00}, but since the basic conclusions are the same, we present
only the NLTE results. A detailed discussion of NLTE vs. LTE models
can be found in \citeone{HRW00}. In the following sections we will
compare the line indices of Balmer lines in these spectra with
those of our observations.

\subsection{The Effect of Temperature}
\label{sec:tempew}

\begin{figure}
  \begin{center}
    \leavevmode
    \epsfig{file=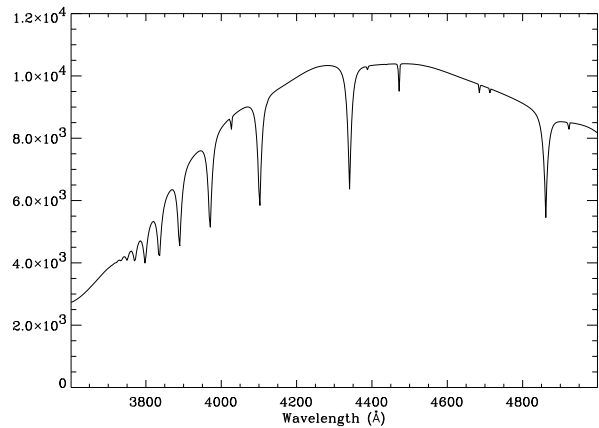,scale=0.47}
    \caption{Model sdB spectrum with $T_{\mathrm{eff}}$=32\,500\,K and
    log\,$g$=5.25, used to determine line index variations. The
    continuum is based on a typical NOT spectrum. Flux is in arbitrary
    units.}
    \label{fig:modspec}
  \end{center}
\end{figure}

For most oscillating stars, it is usually assumed that variations are
predominantly effected by temperature changes, and that radius changes
(and hence surface gravity changes) are negligible \citeeg{KBV95}. In
this case, the oscillation amplitude in line index is related to
the temperature fluctuations by the relation

\begin{equation}
  \label{eq:EWvT}
  \frac{\delta
    \Lambda}{\Lambda}=\frac{\partial\log\Lambda}{\partial\log T}\frac{\delta
    T}{T}.
\end{equation}
\begin{figure}
\center{\epsfig{file=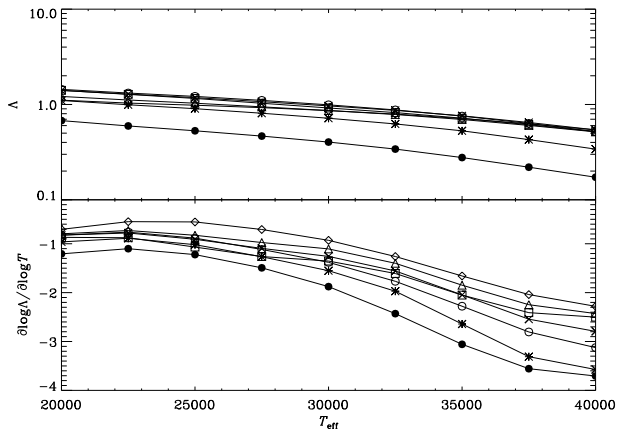,scale=0.47}}
\caption{LTE model line index, $\Lambda$ (\textit{top panel}) and its
  derivative $\partial\log\Lambda/\partial\log T$ (\textit{bottom
  panel}), both as a function of effective temperature. The symbols
  have the following meaning: diamonds -- H$\beta$; triangles --
  H$\gamma$; squares -- H$\delta$; crosses -- H$\epsilon$; open
  circles -- H8; stars -- H9; filled circles -- H10.}
\label{fig:modeltemp}
\end{figure}

For cool stars $\partial\log\Lambda/\partial\log T$ is positive, but for
hot stars such as sdBs, it is expected to be negative. (Note that for
A stars it is approximately zero). To investigate the behaviour of
the line index with varying temperature for each Balmer line, we show
in Figure \ref{fig:modeltemp} both $\Lambda$ and
$\partial\log\Lambda/\partial\log T$ as a functions of
temperature. Here we have used LTE models with a larger range of
effective temperatures and the surface gravity fixed at $\log
g$=5.5. Prior to calculating the line indices, we rebinned the model
spectra to the approximate dispersion of our observed spectra and
multiplied these spectra by a typical continuum from the NOT
observations. We also used exactly the same parameters in our line
index software for both models and observations.

Figure \ref{fig:modeltemp} is similar to Figure 2 of \citeone{BKR96},
which was made for cooler main sequence stars. An important difference
is the scale of change of line index with temperature. For cool main
sequence stars there is a factor of three difference over around
1000\,K for H$\beta$, while for our sdB models, the difference for
H$\beta$ is about a factor of two over 20\,000\,K. The line index of
all Balmer lines except H9 and H10 seems to converge at about
40\,000\,K, although without modelling beyond this temperature we
cannot be certain (and $\sim$40\,000\,K is the upper limit of the sdB
regime anyway). As is expected, the change in line index with
temperature is negative, that is, the line index of hotter sdBs is
smaller.

To investigate the ``pseudo-continuum'' idea suggested in Section
\ref{sec:explain}, we rebinned our NLTE
model spectra to the approximate dispersion of our observed spectra
and again multiplied the resultant spectra by a typical continuum from
the NOT observations. Our goal was to use a model that was as close to the
observations as possible. We calculated line indices for each of our
models, once again using the same parameters as for the observations. To
simulate oscillations, we took the difference between the line index of
each model Balmer line at each of the different temperatures listed
above. The surface gravity of the models was fixed at log\,$g$=5.5. The
change in line index, $\Delta\Lambda$, is approximately linearly
proportional to the change in temperature $\Delta
T_{\mathrm{eff}}$. Because of this relationship, we can scale the
line index ``amplitude'' to a more realistic value of $\Delta
T$. We chose $\Delta T\sim500$\,K, which is an order-of-magnitude
estimate based on photometric amplitudes. Finally, we normalised these
differences by the line index values at 32\,500\,K (the
approximate effective temperature of PG\,1605+072), allowing us to
compare fractional changes directly with our observations.

\begin{figure}
\center{\epsfig{file=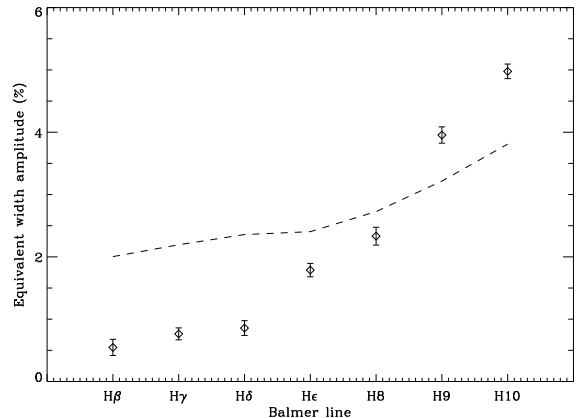,scale=0.47}}
\caption{line index of the highest peak (2742.85\,$\mu$Hz) in
  the 2000 amplitude spectrum compared with a simple model with
  temperature variations of $\sim$500\,K (dashed line).}
\label{fig:ampcomp}
\end{figure}

The model line-index amplitude is plotted as the dashed line in
Figure \ref{fig:ampcomp}. Also shown is the line-index amplitude
of the 2742.85\,$\mu$Hz peak from the March--May 2000 observations. Both
model and observations show an upward trend moving to bluer wavelengths,
but the amplitudes do not match. A similar result is seen for each of
the other frequencies in Table 7 of \pII. Clearly then, these unusual
variations depend on more than just temperature, particularly for the
higher-order Balmer lines. We now consider the
effect of changing surface gravity as well as temperature. 

\subsection{Add Gravity and Stir}
\label{sec:addgrav}

Here we assume that the relationship between change in surface gravity
and change in line index is similar to Equation \ref{eq:EWvT},
and that we can use the simple formula
\begin{equation}
  \label{eq:EWvTandg}
  \frac{\delta
    \Lambda}{\Lambda}=\left( \frac{\delta\Lambda}{\Lambda}
    \right)_{T_\mathrm{eff}}+\left( \frac{\delta\Lambda}{\Lambda} \right)_{\mathrm{log}g}
\end{equation}
to combine the effects of $T_{\mathrm{eff}}$ and log\,$g$. 
The first term on the right of Equation \ref{eq:EWvTandg} is the
line index at fixed $T_{\mathrm{eff}}$, and the second term is
at fixed log\,$g$. This linearity assumes that the oscillations are
adiabatic, which implies that $T_\mathrm{eff}$ and log\,$g$ are in
phase.

\begin{figure}
\center{\epsfig{file=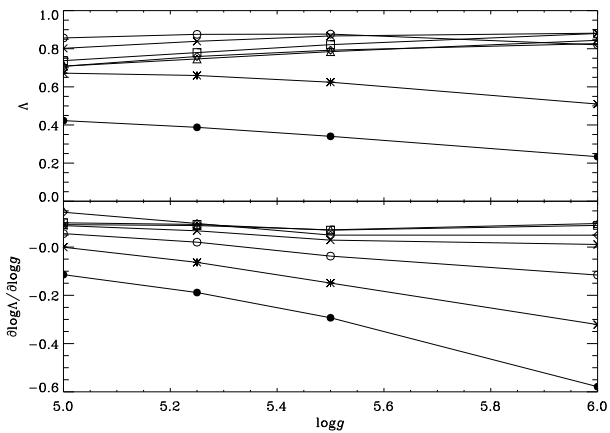,scale=0.47}}
\caption{LTE model line index, $\Lambda$ (\textit{top panel}) and its
  derivative $\partial\log\Lambda/\partial\log g$ (\textit{bottom
  panel}), both as a function of surface gravity. Symbols have the
  same meaning as Figure \ref{fig:modeltemp}.}
\label{fig:modellogg}
\end{figure}

We show $\Lambda$ and $\partial\log\Lambda/\partial\log g$ as
functions of $\log g$ in Figure \ref{fig:modellogg}. The important
thing to note about this figure is that the line index of low-order Balmer
lines stays roughly constant with $\log g$, while for the higher order
lines the line index decreases dramatically. This is related to the
psuedo-continuum effect discussed earlier. The implication of Figures
\ref{fig:modeltemp} and \ref{fig:modellogg} is that line index
amplitudes should be highest for higher order Balmer lines, which is
what we see. We can now try to quantify this effect.

\begin{figure}
  \begin{center}
    \leavevmode
    \epsfig{file=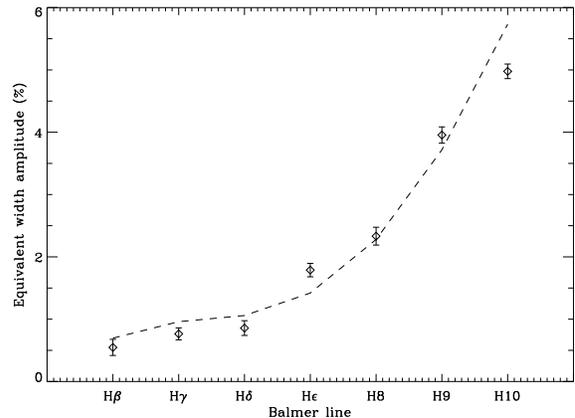,scale=0.47}
    \caption{Balmer line line index with frequency
      2742.85\,$\mu$Hz, as a function of Balmer line, compared with a
      simple model (dashed line). The fit shown here gives $\delta T =
      \sim$372\,K and $\delta\mathrm{log}g = \sim$0.047.}
    \label{fig:2742.85fit}
  \end{center}
\end{figure}

Using a similar method to that used for temperature changes above,
but keeping $T_{\mathrm{eff}}$ fixed at 32\,500\,K, we have
calculated the approximate effect of changing log\,$g$. To fit the
combination of these effects, we minimised the rms scatter of the
difference between observations and the model. The best-fitting
combination of the temperature and surface gravity effects is shown in
Figure \ref{fig:2742.85fit}. The fit appears to be, in general, very
good, although there is some discrepancy at H10. For this frequency,
we find that $\Delta T\sim372$\,K and
$\Delta$log\,$g\sim0.047$. Whether or not these values are reasonable
will be discussed in the next section.

Assuming that this method is valid, we have determined the temperature
and surface gravity changes for each frequency in Table 6 of \pII, as
well as the frequencies measured from the 1999 observations (Table 5
of \pII), and these are shown in Table \ref{tab:dTdg}. The velocity
amplitudes derived in \pII\ are shown in the last column. The
uncertainties in both $\Delta T_{\mathrm{eff}}$ and $\Delta$log\,$g$
were found by fixing one of the $(\delta\Lambda/\Lambda)_i$, and
changing the other until the rms increased by the mean error of the
measured line index amplitudes. We have redone our calculations
including the effects of rotation ($v\sin i=39$\,km\,s$^{-1}$ for
PG\,1605+072; \citebare{HRW99}), and find that the values in Table
\ref{tab:dTdg} remain the same within 0.1\%.

\begin{figure}
  \begin{center}
    \leavevmode
    \epsfig{file=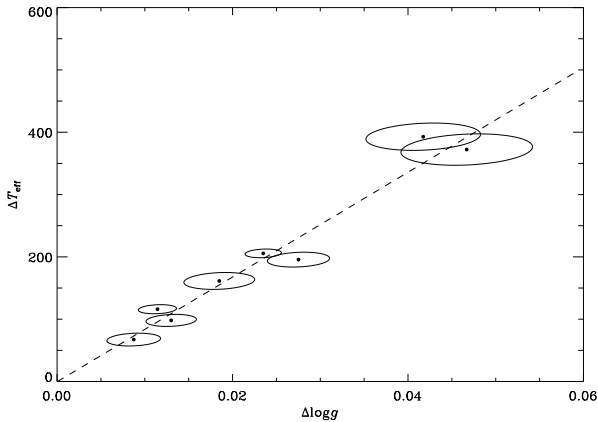,scale=0.47}
    \caption{$\Delta T_{\mathrm{eff}}$ as a function of
    $\Delta\mathrm{log}g$ for the amplitudes shown in the top section
    of Table \ref{tab:dTdg}. The error ellipses were calculated using
    linear regression, and show 1$\sigma$ errors. The dashed line
    represents a weighted least-squares fit to the data.}
    \label{fig:Tvlogg}
  \end{center}
\end{figure}

In Figure \ref{fig:Tvlogg} we show $\Delta T$ as a
function of $\Delta\mathrm{log}g$ for both sets of observations, with
1$\sigma$ error ellipses determined from linear regression. Most of
the points lie on or near the straight line. The slope of the line is
8471\,K\,dex$^{-1}$. This can be compared with future models. If there
is linearity, it would be consistent with all modes having the same
$l$ value. Modelling by \citeone{Kawaler99} found 4 modes to be $l=1$
and one to be $l=2$. We have detected 4 of the 5 modes predicted by
\citename{Kawaler99}, including the mode predicted to be $l=2$ (around
2102\,$\mu$Hz). In Figure \ref{fig:Tvlogg}, this mode appears to lie
on the same line as the other modes. Further modelling is required to
determine whether this is inconsistent with \citename{Kawaler99}'s model. 

\begin{table}
\caption{Changes in temperature and surface gravity for each frequency
      found in \pII\ derived from Balmer line line indices for
      both 1999 and 2000 observations.}
\label{tab:dTdg}
\begin{center}
\begin{tabular}{lcccc}
Year & Frequency & $\Delta T_{\mathrm{eff}}$ & $\Delta$log\,$g$ &
$v_{\mathrm{\pII}}$ \\
& ($\mu$Hz) & (K) & & (km\,s$^{-1}$) \\
\hline
2000 & 2742.85 & \llap{3}72$\pm$24 & 0.047$\pm$0.006 & 7.17 \\
& 2742.47 & \llap{1}96$\pm$13 & 0.028$\pm$0.004 & 4.45 \\
& 2102.83 & \llap{1}61$\pm$15 & 0.018$\pm$0.004 & 3.62 \\
& 2102.48 & \llap{3}92$\pm$24 & 0.042$\pm$0.006 & 8.47 \\
& 2101.57 & \llap{1}16$\pm$ 7 & 0.011$\pm$0.002 & 3.40 \\
& 2075.72 & \llap{2}05$\pm$ 9 & 0.023$\pm$0.002 & 4.27 \\
& 1985.75 & 98$\pm$11 & 0.013$\pm$0.003 & 4.13 \\
& 1891.01 & 67$\pm$11 & 0.009$\pm$0.003 & 1.99 \\
\hline
1999 & 2742.63 & \llap{2}83$\pm$14 & 0.051$\pm$0.004 & 5.80 \\
& 2102.15 & \llap{5}60$\pm$17 & 0.062$\pm$0.005 & \llap{1}1.20 \\
& 2075.29 & \llap{1}94$\pm$ 4 & 0.018$\pm$0.001 & 2.41 \\
& 1890.98 & 54$\pm$16 & 0.003$\pm$0.004 & 3.14 \\
\hline
\end{tabular}
\end{center}
\end{table}

\section{Discussion}
\label{sec:disc-ew}

While the model of line index oscillations we have presented
here is crude, it appears to explain the observations quite well. We
now must consider whether the changes in temperature and surface
gravity we determined are reasonable.

\subsection{Temperature Changes}
\label{sec:tempchange}

We can use the Johnson $B$ photometry presented in \pI\ to test the
plausibility of the temperature changes we have calculated. First, we
need to convert these photometric amplitudes to fractional bolometric
luminosity changes. \citeone{K+B95} gave the following relationship
between the fractional bolometric luminosity variation $(\delta
L/L)_{\mathrm{bol}}$ and the observed fractional luminosity variation
$(\delta L/L)_{\lambda}$

\begin{equation}
  \label{eq:KB95}
\left(\frac{\delta L}{L}\right)_{\mathrm{bol}}=\left(\frac{\delta L}{L}\right)_{\lambda}\frac{\lambda}{\lambda_{\mathrm{bol}}}
\end{equation}
where
\begin{equation}
  \label{eq:lambdabol}
  \lambda_{\mathrm{bol}}=\frac{hc}{kT_{\mathrm{eff}}}.
\end{equation}

Equation \ref{eq:KB95} is a linearised expression, assuming the star
is radiating as a blackbody. We have derived the exact form of this
equation, where we also allow for the radius change in the star (in
PG\,1605+072, assuming radial modes, the radius change is 0.1--0.9\%):

\begin{eqnarray}
\left(\frac{\delta L}{L}\right)_{\mathrm{bol}}& = &
\left[\left(\frac{\delta L}{L}\right)_{\lambda}-2\frac{\delta
R}{R}\right]\times \nonumber \\
& & \hspace{0.5cm} \frac{\lambda}{\lambda_{\mathrm{bol}}}(1-e^{-4\lambda_{\mathrm{bol}}/\lambda})+2\frac{\delta R}{R}.
\label{eq:boloboltzrad}
\end{eqnarray}

As an example, consider the 2102.15\,$\mu$Hz mode in our 1999
data (\pI). The radius change is $\sim$0.9\% and the photometric
amplitude is 3.2\%, implying that $(\delta L/L)_{\mathrm{bol}}\sim$\,10\%.
This leads to a temperature change of $\sim$\,490\,K, which is roughly
consistent with the value of 560\,K shown at the bottom
of Table \ref{tab:dTdg}. The temperature changes from photometry for
the other modes from Table \ref{tab:dTdg} are also about the same
order of magnitude as those determined from line index. We conclude
from this that the temperature values we have derived are the correct
order of magnitude.

\subsection{Surface Gravity Variations}
\label{sec:loggchange}

It is possible that we are measuring
changes in the \textit{effective} surface gravity rather than the true
surface gravity. The effective surface gravity is a combination of the
true surface gravity and other forces acting on the stellar
photosphere. In order to test whether we are measuring true surface
gravity variations, we have calculated the radius change expected from
the best-fit variation in surface gravity at each frequency. These
changes are shown in Table \ref{tab:Rchange}, along with the radius
changes calculated from velocity amplitudes, assuming all modes are
radial. The difference between the radius change for a radial and
non-radial mode is a scaling factor which is dependent on the
inclination angle and limb darkening. For example, with no limb
darkening and with the star pole on, an $l=1, m=0$ mode will have an
apparent radius change $\sim$30\% larger than an $l=0$ mode with the
same frequency, while an $l=2, m=0$ mode will have a radius change
$\sim$10\% smaller \cite{ChD94}. \citeone{BKR96} presented an analysis
for solar-like stars that also considered limb darkening. They found
that the mode sensitivity of the line index of Balmer lines is similar
to that of Balmer-line velocity. 
\begin{table}
\caption{Change in radius determined from log\,$g$ variations,
compared with change in radius expected from velocities for a radial
mode.}
\label{tab:Rchange}
\begin{center}
\begin{tabular}{lccc}
Year & Frequency & $\left( \frac{\Delta R}{R} \right)_{\mathrm{log}g}$ &
$\left( \frac{\Delta R}{R} \right)_{\mathrm{vel}}$ \\
& ($\mu$Hz) & (\%) & (\%) \\
\hline
2000 & 2742.85 & 5.41 & 0.44 \\
& 2742.47 & 3.22 & 0.27 \\
& 2102.83 & 2.07 & 0.29 \\
& 2102.48 & 4.84 & 0.68 \\
& 2101.57 & 1.27 & 0.27 \\
& 2075.72 & 2.65 & 0.35 \\
& 1985.75 & 1.50 & 0.35 \\
& 1891.01 & 1.04 & 0.18 \\
\hline
1999 & 2742.63 & 5.87 & 0.36 \\
& 2102.15 & 7.14 & 0.90 \\
& 2075.29 & 2.07 & 0.20 \\
& 1890.98 & 0.35 & 0.28 \\
\hline
\end{tabular}
\end{center}
\end{table}

All of the values of $\Delta R/R$ calculated from the log\,$g$ changes
are larger than those calculated from velocity by at least a factor of
five. 
A more realistic interpretation of the line index measurements
requires sophisticated models. Phase
dependent synthetic spectra need to be calculated that also account for 
limb darkening and rotation. As a prerequisite for such model spectra,
the time dependent surface distribution of $\mathrm{T}_{eff}$,
${\mathrm{log}}\,g$ and the velocity field have to be determined.
\citeone{BRUCE97} has developed such a program (BRUCE) for the analysis of
pulsations in rotating hot massive stars.
\citeone{Falter2001} recently wrote a code for modelling synthetic
spectra of sdB stars from BRUCE output which shall be used to analyse
the line index variations. A much larger data set has been acquired in
May/June 2002 by spectroscopic and photometric multi-site campaign
which became known as the Multi-site 
spectroscopic telescope (MSST, \citebare{HDS2002}). We will use the more 
detailed modelling when we shall analyse this much larger data set. 

However we may also have to check the validity of the hydrostatic 
approximation. \citeone{JWP01} encountered a similar problem in the 
analysis of the extreme helium star V652\,Her; for this radially pulsating 
star the discrepancy between radii derived from surface gravity and 
spectrophotometry is about a factor of two. They found that at least for 
10\% of the pulsation cycle of this star the assumption of hydrostatic 
equilibrium is not valid, due to shocks moving through the atmosphere. 
For PG\,1605+072 the speed of sound is more than twice as large as for 
V652\,Her due to its different chemical composition. A study of such effects 
are beyond the scope of this paper and should be done after the
more detailed hydrostatic spectral analysis outlined above has been carried 
out.

\section{Conclusions}

We have detected line index variations in the pulsating sdB star
PG\,1605+072. There is a strong dependence of line-index
oscillation amplitude on Balmer line. We have developed a simple model,
assuming hydrostatic equilibrium, where we use model spectra at as
close to the same resolution and sampling of the observations as
possible. Using this model, we have
shown that despite not measuring true equivalent width variations, we
can use the observables to infer effective temperature and effective
surface gravity variations in PG\,1605+072. 

Using more detailed models, we should be able to measure the $\Delta
T$ and $\Delta$log\,$g$ independently of photometry and velocity
measurements. With this information, we will have another set of
amplitudes which we can use for mode identification of sdB stars.

\vspace{0.3cm}
%\acknowledgements

The authors would like to thank Mike Ireland for helpful discussions
on error ellipses, and Steve Kawaler for continuing fruitful
discussions on sdB evolution and interiors. This work was supported
by an Australian Postgraduate Award (SJOT), the Australian Research
Council, the Danish National Science Research Council through its
Center for Ground-based Observational Astronomy, and the Danish
National Research Foundation through its establishment of the
Theoretical Astrophysics Center.

\end{document}